\documentclass[%
 aps,
 prx,
 amsmath,
 amssymb,
 reprint,
 longbibliography
]{revtex4-2}
\usepackage[utf8]{inputenc}
\usepackage{graphicx}
\usepackage{dcolumn}
\usepackage{bm}
\usepackage[T1]{fontenc}
\usepackage{mathptmx}
\usepackage{mathtools}
\usepackage{etoolbox}
\usepackage{siunitx}
\usepackage[draft]{changes}

\usepackage{hyperref}
\usepackage[english]{babel}

\DeclareUnicodeCharacter{2009}{ }
\DeclareUnicodeCharacter{03BC}{$\mu$}
\DeclareUnicodeCharacter{2215}{/}

\makeatletter
\def\@email#1#2{%
 \endgroup
 \patchcmd{\titleblock@produce}
  {\frontmatter@RRAPformat}
  {\frontmatter@RRAPformat{\produce@RRAP{*#1\href{mailto:#2}{#2}}}\frontmatter@RRAPformat}
  {}{}
}

\makeatother
\begin{document}


\title[GE paper - figures phase v1]{An rf-SQUID-based traveling-wave parametric amplifier with -84 dBm input saturation power across more than one octave bandwidth}
\author{Victor Gaydamachenko}
\thanks{These authors contributed equally.}
\affiliation{
 Physikalisch-Technische Bundesanstalt, Bundesallee 100, 38116 Braunschweig, Germany
}
\author{Christoph Kissling}
\thanks{These authors contributed equally.}
\affiliation{
 Physikalisch-Technische Bundesanstalt, Bundesallee 100, 38116 Braunschweig, Germany
}
\author{Lukas Grünhaupt}
\thanks{Corresponding author: lukas.gruenhaupt@ptb.de}
\affiliation{
 Physikalisch-Technische Bundesanstalt, Bundesallee 100, 38116 Braunschweig, Germany
}

\date{\today}

\begin{abstract}
Traveling-wave parametric amplifiers (TWPAs) have become an essential tool for the readout of quantum circuits and the search for dark matter. We report on the implementation of an rf-SQUID-based Josephson TWPA with an average saturation power of \qty{-84}{dBm}, while providing an average power gain of \qty{20}{\decibel} from \qtyrange[range-units = single]{3.5}{8.5}{\giga\hertz}. 
This wide bandwidth is enabled by reducing the curvature of the dispersion in the signal band while suppressing detrimental mixing processes.
With 2393 rf-SQUIDs our device has a comparable number of nonlinear elements, but achieves ten times higher saturation power than previous works, showing that using rf-SQUIDs results in a more favorable trade-off between device length and saturation power.
Harnessing wideband characterization techniques we determine the TWPA's excess noise above the quantum limit to be \numrange[]{0.8}{1.5} photons.
In addition, we validate signal-to-noise ratio improvement as a practical measure for tuning the TWPA to minimize the total system noise, and demonstrate that its optimum does not coincide with the highest gain.
\end{abstract}

\maketitle

\section{Introduction}

Over recent years, near-quantum-limited parametric amplifiers have gained widespread use across various fields of scientific research. By significantly improving the signal-to-noise ratio of experiments involving microwave signals with powers as low as a few femtowatts, they have advanced the technical capabilities of experimental setups. 
Such amplifiers have demonstrated exceptional performance in experiments targeting dark matter detection \cite{Brubaker2017, Backes2021, DiVora2023, Bartram2023, Jiang2023, Uchaikin2024}, reading out signals from microwave SQUID multiplexers \cite{Malnou2023}, and enhancing the sensitivity of microwave single-photon detectors \cite{Zobrist2019, Mohammad2024} or phonon detectors \cite{Ramanathan2024}. 
Moreover, parametric amplifiers have become indispensable tools in the fields of quantum sensing and quantum information \cite{Vijay2011, Bultink2018, White2023, Elhomsy2023, Wang2023, CastellanosBeltran2025}, where their ability to amplify weak signals with minimal added noise \cite{Caves1982} is crucial.

Parametric amplifiers can be divided into two groups: resonator-based parametric amplifiers \cite{CastellanosBeltran2007, Eichler2014, Winkel2020, Sivak2020, White2023} and traveling-wave parametric amplifiers (TWPAs) \cite{Macklin2015, Planat2020, Ranadive2022, Perelshtein2022, Qiu2023, White2015, Miano2019, FadaviRoudsari2023, Nilsson2024, Simbierowicz2021, Bockstiegel2014, HoEom2012, Vissers2016, Zobrist2019, Malnou2021, Klimovich2023, Faverzani2024, Adamyan2016, Chaudhuri2017, Ranzani2018, Goldstein2020, Kern2023, Giachero2024}. The former generally exhibit lower noise, while the latter typically provide bandwidths of \qty{1}{\giga\hertz} or more. This comes at the cost of increased circuit complexity, such as sophisticated dispersion engineering to enable phase-matching and suppress unwanted frequency mixing. 
The wider bandwidth enables greater capacity for frequency multiplexing in applications such as qubit or detector readout.
This feature makes TWPAs particularly advantageous for scaling up quantum computers, where simultaneous readout of many qubits is necessary. Their number is further constrained by the saturation power of the amplifier chain.
Future qubit designs resilient to higher readout powers might necessitate amplifiers with higher power handling capability \cite{Gusenkova2021}. 

Josephson traveling-wave parametric amplifiers (JTWPAs) typically have a saturation power on the order of \qty{-100}{dBm} at a gain of \qty{20}{\decibel} \cite{Macklin2015, Planat2020, Ranadive2022, Perelshtein2022, Qiu2023}.
TWPAs based on high-kinetic-inductance films (KI-TWPAs) \cite{HoEom2012, Vissers2016, Zobrist2019, Malnou2021, Klimovich2023, Faverzani2024} achieve higher saturation power (\qtyrange[range-units = single]{-63}{-47}{dBm}), but operate at pump powers \numrange[range-phrase = --]{4}{5} orders of magnitude higher than those of JTWPAs.
Such strong pump tones require more isolation to shield qubits, and pump cancellation techniques to avoid saturation of the next amplification stages, both adding extra complexity to the experimental setup \cite{Esposito2021}. 

Further, intermodulation distortion must be considered when more than one signal is amplified. 
It arises from undesired mixing processes between the signals and increases drastically as the saturation power of the amplifier is approached. 
This effect can result in significant crosstalk between amplified signals, potentially increasing readout errors within the system \cite{Remm2023}. 
A proposed mitigation strategy is to use a large detuning between the pump and signal frequencies and to increase the saturation power of the TWPA. 

To address these issues, we have designed and fabricated a TWPA based on rf-SQUIDs \cite{Zorin2016}.
It operates in the three-wave-mixing (3WM) regime, resulting in a large separation between pump and signal frequencies.
The dispersive and nonlinear properties are designed to achieve simultaneously a wide bandwidth and high saturation power.
We engineer the dispersion harnessing a multiperiodic variation of the capacitances to ground, enabling simultaneous phase matching of parametric amplification while suppressing unwanted mixing products \cite{Gaydamachenko2022}.
In contrast to other 3WM-TWPAs, which typically rely on a low plasma frequency to achieve the latter \cite{Malnou2021, Perelshtein2022}, our approach allows lower dispersion in the signal band, thereby increasing the attainable amplification bandwidth \cite{Malnou2021}. 
We demonstrate an average gain of \qty{20}{\decibel} between \qtylist[list-units = single]{3.5; 8.5}{\giga \hertz}, which exceeds one octave and is, to the best of our knowledge, the widest relative bandwidth achieved in TWPAs. 

Furthermore, our device achieves a saturation power of \qty{-84}{dBm}, more than one order of magnitude higher than reported for other JTWPAs. 
To achieve this, we leverage the fact that the linear and nonlinear properties of rf-SQUIDs can be effectively decoupled.
Hence, we design the rf-SQUIDs to be weakly nonlinear, aiming for higher pump powers, consequently leading to proportionally increased saturation power. 
In this way we break the trade-off between saturation power and device length ruling the design of JTWPAs based on Josephson junctions or dc-SQUIDs \cite{OBrien2014}.
Indeed, with 2393 rf-SQUIDs our device has a comparable number of nonlinear elements to those reported in Refs.~\cite{Macklin2015, Qiu2023, Planat2020}, but achieves a tenfold increase in saturation power.

\section{Device design and fabrication}
\begin{figure}
    \includegraphics[width=8.6cm]{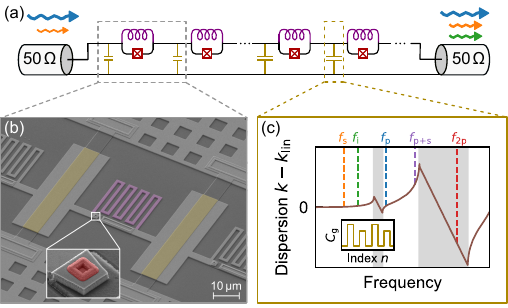}
    \caption{Circuit diagram of the rf-SQUID-based TWPA, experimental implementation, and dispersion relation. 
    (a) A strong pump tone (blue arrow) with frequency $f_\mathrm{p}$ and a weak signal tone (orange arrow) with frequency $f_\mathrm{s}$  propagate along the amplifier and interact via the nonlinearity of the rf-SQUIDs, resulting in energy transfer from the pump to the signal and creation of an idler tone (green arrow) at frequency $f_\mathrm{i} = f_\mathrm{p} - f_\mathrm{s}$. The rf-SQUIDs are formed by a Josephson junction (red) shunted by a linear inductor (violet). We periodically vary the capacitance to ground $C_\mathrm{g}$ (yellow) to realize dispersion engineering [see panel (c)].
    (b) False-colored scanning electron microscopy image of an rf-SQUID between two capacitors to ground [colors according to panel (a)]. Zoom-in shows a junction of size $1 \times \qty{1}{\micro \meter \squared}$. 
    (c) Sketched dispersion relation, obtained using multiperiodic capacitance variation (mPCV), which results in two stopbands (gray shaded regions) \cite{Gaydamachenko2022}. 
    The second stopband mismatches unwanted mixing processes, such as $f_\mathrm{p+s} 
    = f_\mathrm{p} + f_\mathrm{s}$ (violet), $f_\mathrm{p+i} = f_\mathrm{p} + f_\mathrm{i}$ (not shown), and second harmonic generation $f_\mathrm{2p} = 2 f_\mathrm{p}$ (red), while the first stopband enables phase-matching between $f_\mathrm{p}$, $f_\mathrm{s}$, and $f_\mathrm{i}$ to allow amplification. Inset shows mPCV scheme using three values for $C_\mathrm{g}$ (see main text).}
    \label{fig-circuit-diagram}
    \label{fig-sem}
\end{figure}

Figure~\ref{fig-circuit-diagram}(a) shows the circuit diagram of the rf-SQUID based TWPA.
The device comprises a chain of $N=2393$ unit cells, each consisting of an rf-SQUID and a capacitor to ground $C_\mathrm{g}$.
Each rf-SQUID, formed by a meandering inductor $L_\mathrm{m}$ shunted by a Josephson junction with critical current $I_\mathrm{c}$ and junction capacitance $C_\mathrm{J}$, is nominally identical and nonhysteretic, i.e., the screening parameter is $\beta_L = L_\mathrm{m}I_\mathrm{c}/\varphi_0 < 1$ \cite{Clarke2006}. Here $\varphi_0 = \Phi_0 / 2 \pi = \hbar / 2e $ is the reduced magnetic flux quantum. 

Considering an injected ac current with a small amplitude $|i| \ll \Phi_0 / L_\mathrm{m}$, and the corresponding phase oscillation $\phi$, we obtain the current-phase relation for the rf-SQUID:
\begin{equation}
    i(\phi) = \frac{\Phi_0}{2\pi L_\mathrm{SQ}(\phi_\mathrm{dc})}[\phi - \beta \phi^2 - \gamma \phi^3 + \ldots]
\end{equation}
\cite{Zorin2016}, where $\phi_\mathrm{dc}$ is the solution of the transcendental equation $\phi_\mathrm{dc}=\Phi_\mathrm{ext} / \varphi_0 - \beta_L \sin \phi_\mathrm{dc}$ for an external magnetic flux bias $\Phi_\mathrm{ext}$ \cite{Likharev_Book_2016}.
Here, $\beta$ and $\gamma$ are the coefficients of second-order and third-order (Kerr) nonlinearity, 
\begin{align}
    \beta = \frac{\beta_L}{2} \frac{\sin \phi_\mathrm{dc}}{(1 + \beta_L \cos\phi_\mathrm{dc})}, \label{eq:beta} \\
    \gamma = \frac{\beta_L}{6} \frac{\cos \phi_\mathrm{dc}}{(1 + \beta_L \cos\phi_\mathrm{dc})}. \label{eq:gamma}
\end{align}
To operate in the 3WM regime while minimizing unwanted Kerr effects, such as cross-phase modulation (XPM) and self-phase modulation (SPM) \cite{Agrawal2005}, we flux bias our device near the point $\phi_\mathrm{dc} \approx \pi / 2$.
The flux-dependence of the rf-SQUID inductance is given by
\begin{equation}
    L_\mathrm{SQ}(\phi_\mathrm{dc}) = L_\mathrm{m} / (1 + \beta_L \cos\phi_\mathrm{dc})
    \label{eq:Lsq}
\end{equation}
and renders the characteristic impedance and the wavenumber of the transmission line flux-dependent. 
For frequencies much lower than the cutoff frequency, they can be expressed as $Z(\phi_\mathrm{dc}) \approx \sqrt{L_\mathrm{SQ}(\phi_\mathrm{dc})/ C_\mathrm{g}}$ and $k_\mathrm{lin}(\phi_\mathrm{dc}) \approx \omega a^{-1} \sqrt{L_\mathrm{SQ}(\phi_\mathrm{dc}) C_\mathrm{g}}$, respectively. 
Here, $\omega=2\pi f$ is the angular frequency and $a$ is the length of the unit cell.

The 3WM process is governed by the energy conservation law $f_\mathrm{p} = f_\mathrm{s} + f_\mathrm{i}$ and the phase matching condition $k_\mathrm{p} - k_\mathrm{s} - k_\mathrm{i} \ll \pi / N$. Here $f_j$ and $k_j$, $j \in \{\mathrm{p}, \mathrm{s}, \mathrm{i}\}$ are the frequencies and wavenumbers of pump, signal, and idler, respectively. In contrast to four-wave-mixing (4WM), in a dispersion-free medium ($k_\mathrm{lin} \propto \omega$), 3WM is naturally phase matched over a wide frequency range.
However, this also leads to the phase matching of unwanted mixing processes, such as up-conversion (e.g., $f_\mathrm{p+s} = f_\mathrm{p} + f_\mathrm{s}$) and second-harmonic generation ($f_\mathrm{2p} = 2f_\mathrm{p}$), limiting the gain \cite{Dixon2020}.
To address this, the primary objective of dispersion engineering in 3WM TWPAs is to suppress unwanted mixing processes while maintaining phase matching for the amplification \cite{Gaydamachenko2022, RenbergNilsson2023}.

Here, we realize this by a multiperiodic variation of the ground capacitances $C_\mathrm{g}$.
By varying the ground capacitance between three values [see inset Fig.~\ref{fig-circuit-diagram}(c)], we create two stopbands as can be seen in Fig.~\ref{fig-circuit-diagram}(c) \cite{Gaydamachenko2022}. 
Phase matching for parametric amplification is achieved by placing the pump frequency slightly above the first stopband, while undesired tones are suppressed due to large phase mismatches, or cannot propagate in the stopband.
Note that the dispersion for frequencies below the first stopband is flat, while above it increases sharply. This allows to fulfill the phase matching condition for a wider frequency range compared to approaches which rely on a lowered plasma frequency to mismatch unwanted processes \cite{Malnou2021, Zorin2021, Perelshtein2022, FadaviRoudsari2023, Nilsson2024}. 

To estimate what saturation power can be attained with the TWPA, we assume a simplified case with perfectly phase-matched pure 3WM parametric amplification, and suppression of unwanted processes. Then, the signal power gain in the center of the signal band, $f_\mathrm{s}\approx f_\mathrm{p}/2$, can be expressed as 
\begin{equation}
    G = \cosh^2\left(|\beta| k_\mathrm{p} \phi_\mathrm{p} Na / 4 \right),
    \label{eq:gain}
\end{equation}
where $\phi_\mathrm{p} = L_\mathrm{SQ} I_p/\varphi_0$ denotes the amplitude of the pump phase oscillation associated with the pump current $I_\mathrm{p}$ \cite{Zorin2016, Kissling_PhD_2025}. 
The incident pump power is $P_\mathrm{p} \approx 0.5 I_\mathrm{p}^2 Z$. 
Next, we suppose that pump depletion is the only cause for gain saturation. 
The input-referred saturation power, conventionally stated as the 1-dB compression point, is approximately found by
\begin{equation}
    P_{\mathrm{1dB}} \approx \frac{P_\mathrm{p}}{4G_0},
    \label{eq:Pssat_3WM}
\end{equation}
or $P_\mathrm{1dB}^\mathrm{\,dBm} \approx   P_\mathrm{p,in}^\mathrm{\,dBm} -  G^\mathrm{\,dB} - 6\mathrm{~dB}$ (see Appendix~\ref{appendix:saturation}). 
In a 4WM-TWPA the Kerr nonlinearity both facilitates parametric amplification and causes SPM and XPM. In this case  
\begin{equation}
    P_{\mathrm{1dB}} \approx \frac{P_\mathrm{p}}{8G_0}
    \label{eq:Pssat_4WM}
\end{equation}
is smaller by a factor of two due to the additional signal-power-induced phase mismatch by SPM and XPM \cite{Kylemark2006}. 

\begin{figure}
    \includegraphics[width=8.6cm]{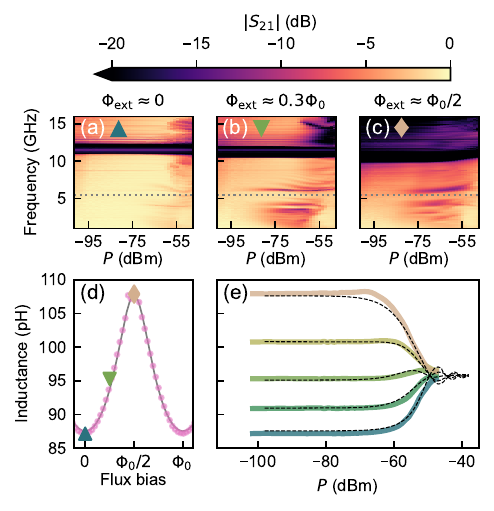}
    \caption{Power dependence of transmission and rf-SQUID inductance. (a-c) Power dependence of $|S_{21}|$ measured at three relevant flux bias points. The stopband position (dark region near \qty{11}{\giga \hertz}) changes as a function of flux due to a change in rf-SQUID inductance [see panel (d)]. At higher powers, the stopband shows a shift caused by self-phase modulation of the probe tone. The direction of the shift reflects the Kerr-nonlinearity coefficient, which is positive (a), negligible (b), and negative (c). 
    For powers larger than \qty{-80}{dBm} in (b) we observe lower transmission between frequencies 1 and \qty{6}{\giga \hertz}, which we attribute to second harmonic generation.
    (d) Cell inductance $L_{\mathrm{cell}}$ as a function of flux bias. Pink points show $L_\mathrm{cell}$ extracted from the experimental data. Gray curve is a fit to the model of the TWPA unit cell (see Appendix~\ref{appendix:non-ideal-squid}). (e) Power dependence of the rf-SQUID inductance, extracted from the phase of $S_{21}$ at \qty{5.5}{\giga \hertz} [gray dotted lines in panels (a-c)]. Dashed lines represent WRspice simulations of the TWPA unit cell.}
    \label{fig-s21-vs-power}
\end{figure}

Since the saturation power is proportional to the pump power, it is limited by the nonlinear element. 
In a Josephson-junction-based TWPA this limit is set by the critical current. 
However, since increasing the critical current decreases the inductance and nonlinearity, for the same gain proportionally more junctions are needed \cite{OBrien2014}.
This establishes a trade-off between saturation power and device length. 
For an rf-SQUID-based TWPA this trade-off is more favorable for two reasons. 
First, the rf-SQUID has two degrees of freedom, which allow to design the nonlinear properties [Eqs.~(\ref{eq:beta}),~(\ref{eq:gamma})] independent from the linear one [Eq.~(\ref{eq:Lsq})], decoupling $k_\mathrm{p}$ from $\beta$ in Eq.~(\ref{eq:gain}). 
Second, thanks to the superconducting shunt $L_\mathrm{m}$ there is no hard limit for the pump current where rf-SQUIDs switch to the voltage state (see Fig.~\ref{fig-s21-vs-power}). 
This allows pumping the TWPA with high amplitudes, limited only by the onset of detrimental effects caused by higher-order nonlinearities. 

Figure~\ref{fig-sem}(b) shows a scanning electron microscope (SEM) image of an rf-SQUID between two capacitors to ground. The device is fabricated in Nb trilayer technology using SIS junctions (see Appendix~\ref{appendix:fab}), on an intrinsic Si wafer with \qty{200}{\nano \meter} of thermal $\mathrm{SiO_2}$. 
The Josephson junctions have a critical current of $I_\mathrm{c} = \qty{0.9}{\uA}$. 
Together with the planned shunting inductance $L_\mathrm{m}=\qty{60}{\pH}$ this yields $\beta_L \approx 0.16$.
The capacitors to ground are realized through plate capacitors using \qty{240}{\nano \meter} thick $\mathrm{SiO_x}$ as the dielectric, and are designed to have the three values $C_{1}=\qty{10.5}{\fF}$, $C_{2}=\qty{68.2}{\fF}$, $C_{3}=\qty{50.4}{\fF}$. 
Groups of six unit cells each have identical capacitances to ground, resulting in a periodicity of 24 unit cells.

\section{Characterization}

We characterize the device in a commercial dilution refrigerator at $\sim\qty{20}{\milli \K}$. Our experimental setup has a pair of coaxial switches to perform \emph{in situ} forward calibration with a female-female SMA adapter as the reference. All cables used between the ports of the switches are nominally of the same length.
To set the operation point we apply a dc current $I_\mathrm{dc}$ to the central conductor of the TWPA transmission line via two bias tees, resulting in an external flux $\Phi_\mathrm{ext} = L_\mathrm{m} I_\mathrm{dc}$ in the rf-SQUID loops.
For details on the experimental setup see Appendix \ref{appendix:setup}. 

\subsection{Pump off measurements}

Figure~\ref{fig-s21-vs-power}(a-c) shows the transmission of the TWPA versus probe tone power for three flux bias points. 
The dark regions around \qty{11}{\giga\hertz} indicate the first stopband, the position of which varies depending on the flux bias. 
At the flux bias points $\Phi_\mathrm{ext} \approx 0$ and $\Phi_0 / 2$ the Kerr-nonlinearity coefficient $\gamma$ has its extrema and results in a power-dependent stopband position due to SPM of the probe tone.
While this is clearly visible for panel (c), in panel (a) it is not obvious.
At $\Phi_\mathrm{ext}\approx  0.3\Phi_0$ the Kerr nonlinearity is negligible, and no power-dependent shift of the stopband is observed. The second-order nonlinearity coefficient $\beta$ is close to its maximum and causes second harmonic generation. This manifests in transmission loss, visible in the darker region for powers ranging from \qtyrange[range-units = single]{-80}{-50}{dBm} across frequencies of \qtyrange[range-phrase = --, range-units = single]{1}{6}{\giga\hertz}.
Similarly, the lossy regions in panel (c) are attributed to third harmonic generation, as well as residual second harmonic generation.
Notably, near the region denoted by the dotted line, this process is suppressed, as the second harmonic falls into the stopband.

\begin{figure*}
    \includegraphics[width=17.8cm]{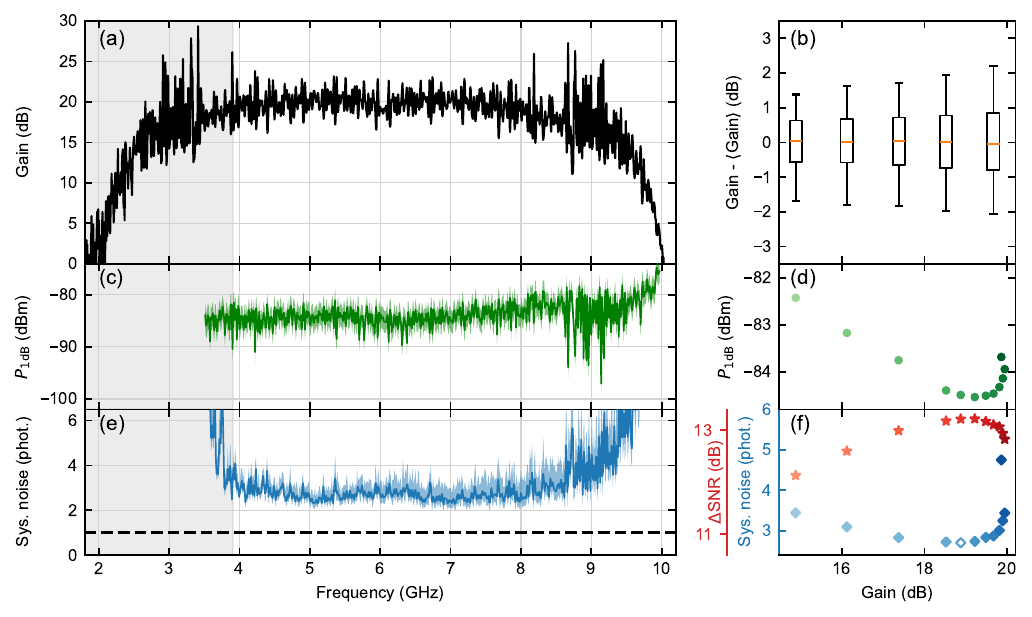}
    \caption{Gain profile, saturation power, and system noise.
    (a) We demonstrate an average gain of \qty{20}{\decibel} from \qtyrange[range-units = single]{4}{8}{\giga\hertz} relative to a through reference. The gray shaded region indicates frequencies where the isolator is out of band (see Appendix~\ref{appendix:setup}). 
    (b) Boxplots showing amplifier gain distribution for different mean gain values for frequencies from \qtyrange[range-units = single]{4}{8}{\giga\hertz}. Whiskers (boxes) represent the 5th (25th) and 95th (75th) percentiles. Orange line indicates the median value. 
    (c) TWPA input saturation power (1-dB compression point $P_\mathrm{1dB}$) as a function of frequency for the gain profile shown in panel (a). 
    Our device reaches an average saturation power of \qty{-84}{dBm}.
    The shaded region indicates the uncertainty of the power calibration. 
    (d) Dependence of average $P_\mathrm{1dB}$ on average gain across the \qtyrange[range-phrase = --, range-units = single]{4}{8}{\giga\hertz} frequency range. Darker color denotes higher pump power. Higher pump power increases the gain, resulting in a decrease of saturation power until the TWPA reaches its maximum gain. A further increase of pump power increases saturation power, but also the total system noise [see panel (f)].
    (e) Using the TWPA as the first amplifier in the readout chain yields a total system noise between 2 and 4 photons at the gain value optimized for lowest average system noise [\qty{19}{\decibel}, see empty marker in panel (f)].
    The shaded region indicates uncertainty of the noise evaluation (see main text).
    (f) Dependence of average total system noise (blue diamonds) and signal-to-noise ratio improvement $\Delta \mathrm{SNR}$ (red stars) on average power gain across the \qtyrange[range-units=single, range-phrase=--]{4}{8}{\giga \hertz} frequency range. Darker color denotes higher pump power. The minimal total system noise (and maximal $\Delta \mathrm{SNR}$) is observed at an average gain of \qty{19}{\decibel}.
    }
    \label{fig-gain_profile}
    \label{fig-system_noise}
\end{figure*}

Figure.~\ref{fig-s21-vs-power}(d) shows the rf-SQUID inductance as a function of external flux. 
By fitting a cascaded transfer matrices model (CTM) \cite{Pozar_book} to the flux-dependent $S_{21}$ spectra for a signal power of \qty{-100}{dBm}, we obtain the inductance values for different external flux values (pink points).
Compared to the simplified circuit diagram [see Fig.~\ref{fig-circuit-diagram}(a)], the practical implementation of the rf-SQUID TWPA unit cell has additional elements, e.g., the inductance of the line connecting two neighboring rf-SQUIDs, which have to be accounted for. 
For our model of the flux-dependent rf-SQUID inductance we take these additional elements into consideration to extract $I_\mathrm{c}=\qty{0.93}{\micro\ampere}$ and $L_\mathrm{m} = \qty{58.6}{\pH}$ (see Appendix~\ref{appendix:non-ideal-squid}). 

Figure~\ref{fig-s21-vs-power}(e) shows the comparison of the experimentally determined power dependence of the cell inductance $L_\mathrm{cell}$ at different flux bias points with that of WRspice simulations \cite{wrspice}.
Note that the rf-SQUID inductance remains almost constant even for high power levels above \qty{-60}{dBm} for $\Phi_\mathrm{ext} \approx 0.3 \Phi_0$, allowing for high pump powers without detrimental SPM and XPM effects.
To take advantage of this rf-SQUID feature we operate our TWPA close to this setpoint. 
In comparison, a Josephson junction with the same inductance of \qty{95}{\pH} switches to the voltage state already at a power of \qty{-65}{dBm}.
The cell inductance is extracted as a function of power from experimental data by analyzing the power-dependent phase of $S_{21}$ at a fixed frequency of $\qty{5.5}{\giga \hertz} \approx f_\mathrm{stopband} / 2$, where the power leakage to the second harmonic is suppressed because of the stopband (see Appendix~\ref{appendix:L_cell_vs_power}).
In WRspice we simulate the behavior of a single unit cell of our TWPA according to the model described above and also extract the power dependent inductance for different flux bias points (see Appendix~\ref{appendix:L_cell_vs_power}). 
Finally, we align the experimental data to the simulation results, which provides an input power calibration at $\qty{5.5}{\giga \hertz}$ and yields an input line attenuation of $\qty{69}{\decibel}$. 

\subsection{Pump on measurements}
\sisetup{uncertainty-mode = separate}

To tune up the TWPA, we vary the flux bias, the pump power and the pump frequency and measure the transmission compared to a through reference. 
Figure~\ref{fig-gain_profile}(a) shows the optimized power gain as a function of signal frequency using the flux bias point $\Phi_\mathrm{ext} \approx 0.33 \, \Phi_0$, pump frequency $f_\mathrm{p}=\qty{12.08}{\giga\hertz}$, and pump power $P_\mathrm{p} \approx \qty{-56}{dBm}$.
To measure the gain we use a signal power $P_\mathrm{s} \approx \qty{-126}{dBm}$ to avoid saturating losses. 
Therefore, we get a lower bound on the gain, which increases by up to \qty{2}{\decibel} for higher signal powers (see Appendix \ref{appendix:tls}).
Using the optimized pump parameters, \qty{90}{\percent} of gain values in the \qtyrange[range-phrase = --, range-units = single]{4}{8}{\giga\hertz} band, sampled with \qty{5}{\mega \hertz} frequency spacing, fall within the range of \qty[separate-uncertainty-units = bracket]{19.8(2.4)}{\decibel}. 
As shown in Figure~\ref{fig-gain_profile}(b), lower gain results in lower ripple.

Gain ripple in TWPAs originate in the interference of multireflections of the signal and idler waves.
This effect can be enhanced by backward amplification due to a reflected pump wave. 
Therefore, to decrease ripple, improved impedance matching is required for both the signal band and the pump frequency.
We investigate the effect of the components following the TWPA by placing a \qty{10}{\decibel} attenuator at its output, which decreases gain ripple, resulting in a gain of \qty[separate-uncertainty-units = bracket]{19.8(1.5)}{\decibel} in the \qtyrange[range-units = single]{4}{8}{\giga \hertz} band. 
All the sampled values between \qtylist[list-units = single]{3.6; 8.3}{\giga \hertz} exceed \qty{17}{\decibel} of gain (see Appendix \ref{appendix:effect-on-ripples}), emphasizing the importance of well-matched external circuitry. 

We also characterize the phase-sensitive gain of the TWPA by measuring the signal gain at $f_\mathrm{s}=f_\mathrm{p} / 2$ as a function of the phase difference between the signal and pump waves.
The maximum ratio between amplification and deamplification, also called phase-sensitive extinction ratio, equals \qty{58}{\decibel} (see Appendix~\ref{app:phase-sensitive-gain}). 

Figure~\ref{fig-gain_profile}(c) illustrates the power-handling capability of our TWPA by showing the input-referred 1-dB compression point $P_\mathrm{1dB}$.
Across the \qtyrange[range-units = single]{4}{8}{\giga \hertz} range, the device achieves an average saturation power of \qty[separate-uncertainty-units = bracket]{-84(2)}{dBm}.
To calibrate the absolute power at the TWPA input across the whole signal band we use a wideband technique (see Appendix~\ref{appendix:powercal_from_noise}), which yields an average attenuation of \qty[separate-uncertainty-units = bracket]{66(2)}{\decibel} in the line, close to the previously quoted attenuation of \qty{69}{\decibel} at \qty{5.5}{\giga \hertz}. 
Using $P_\mathrm{p} \approx \qty{-56}{dBm}$ calibrated in the same way, Eq.~(\ref{eq:Pssat_3WM}) predicts with \qty{-82}{dBm}, in good agreement with the experimental value.
Fig.~\ref{fig-gain_profile}(d) illustrates the relationship between the average saturation power and the average gain. Increasing the pump power results in higher gain, but simultaneously reduces the saturation power until the TWPA reaches its maximum gain. Increasing the pump power further does not lead to significantly higher gain, but a slight rise of $P_\mathrm{1dB}$.

\section{Noise performance}
\sisetup{
    uncertainty-mode = compact}

The noise is characterized with a modified variant of the commonly used Y-factor method. A variable-temperature \qty{50}{\ohm} load terminating a superconducting coaxial cable connected to the TWPA input serves as a calibrated noise source (see Appendix~\ref{appendix:setup}). 
For noise modeling and evaluation of the measured data, we use a two-mode model following the approach described in Refs.~\cite{Malnou2021, Malnou2024}.
To minimize a systematic error in the evaluation, TWPAs should, in principle, be modeled as multi-mode amplifiers. Such a model not only accounts for the conversion of idler input noise to signal output noise, but also for the noise at sideband frequencies \cite{Peng2022, Kissling_PhD_2025}. 
However, experimentally quantifying the contribution of the sidebands is challenging, because it requires absolute power calibration at the TWPA input for frequencies above the pump frequency.
Here we neglect the impact of sidebands on the output noise due to their strong suppression thanks to the multiperiodic capacitance variation (see Appendix~\ref{appendix:sb_excess_noise}).

The two-mode noise model of the amplification chain used for fitting the experimental data is
\begin{equation}
    N_\mathrm{out}^\mathrm{s}(T) = G_\mathrm{sys}\left(N_\mathrm{in}^\mathrm{s}(T)+ \frac{G_\mathrm{si}}{G_\mathrm{ss}}N_\mathrm{in}^\mathrm{i}(T) + N_\mathrm{sys, exc}\right),
    \label{eq:noise_power}
\end{equation}
where the output noise is expressed as a number of photons $N_\mathrm{out} = P_\mathrm{out} / h f_\mathrm{s} B$ at frequency $f_\mathrm{s}$, and $P_\mathrm{out}$ is the noise power measured with a spectrum analyzer (SA) in the resolution bandwidth $B$.
By varying the temperature $T$ of the load, we change its Johnson-Nyquist noise $N_\mathrm{in}^\mathrm{s, i}= 0.5 \coth (hf_\mathrm{s, i} / 2 k_\mathrm{B} T)$, where $k_\mathrm{B}$ is the Boltzmann constant. Thus, we vary the noise incident to the TWPA at both the signal and idler frequencies. 
The conversion gain of idler input noise to the signal output noise is expressed by  $G_\mathrm{si}$, which is ideally equal to the signal power gain $G_\mathrm{ss}\gg 1$. 
Fitting Eq.~(\ref{eq:noise_power}) to the experimental data ($P_\mathrm{out}$) yields the total system gain $G_\mathrm{sys}$ and system excess noise $N_\mathrm{sys, exc}$, the latter of which describes the noise above the standard quantum limit of one photon. 

Figure~\ref{fig-gain_profile}(e) shows the total system noise of our readout chain using the TWPA as the first amplifier, setting the gain asymmetry $G_\mathrm{si} / G_\mathrm{ss} = 1$ [see Eq.~(\ref{eq:noise_power})]. 
Within the \qtyrange[range-units = single]{4}{8}{\giga \hertz} range, the total system noise is between $2.2$ and $3.8$ photons. 
To estimate the uncertainty of the system noise (blue shaded area) we take into account a power level uncertainty of \qty{0.1}{\decibel} of the SA, a temperature stability of $\pm \qty{3}{\milli \K}$ on top of the uncertainty provided by the calibration of the thermometer, and a gain asymmetry of $\pm \qty{1}{\decibel}$.

Figure~\ref{fig-system_noise}(f) shows how the average system noise and signal-to-noise-ratio improvement $\Delta \mathrm{SNR} = \mathrm{SNR}_\mathrm{pump\,on} / \mathrm{SNR}_\mathrm{pump\,off}$ change with the average TWPA gain. 
The operational point with the lowest system noise corresponds to the maximum $\Delta \mathrm{SNR}$, but does not align with the TWPA's maximum gain. 
Additionally, we compare the $\Delta \mathrm{SNR}$ calculated using the results of the Y-factor measurement to the directly measured $\Delta \mathrm{SNR}$ and find a root mean square (RMS) deviation of less than \qty{0.5}{\decibel} (see Appendix~\ref{appndix:snr-from-y-factor}).
While $\Delta \mathrm{SNR}$ is a relative measure and does neither quantify the system noise nor the TWPA's added noise, it serves as a practical metric for optimizing the TWPA's operation point, especially in measurement setups where no calibrated noise source is available.
Therefore, for practical applications, \mbox{TWPAs} should be tuned for lowest noise performance rather than targeting the highest possible gain.

\begin{figure}
    \includegraphics[width=8.6cm]{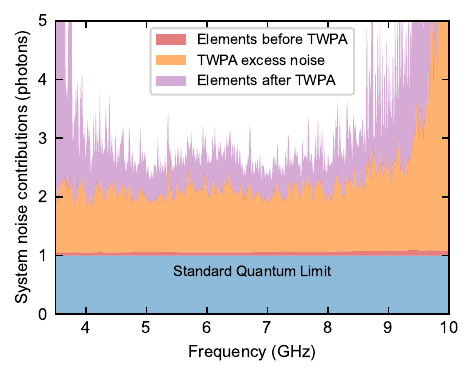}
    \caption{Contributions to the total system noise as a function of frequency for a TWPA gain of \qty{19}{\decibel}. The TWPA excess noise is between 0.8 and 1.5 added photons, which we attribute to dielectric loss in the $\mathrm{SiO_x}$ of plate capacitors to ground \cite{Macklin2015} and the generation of sidebands \cite{Peng2022}.
    Added noise of the elements after the TWPA contributes 0.6 photons on average. The noise contribution of elements before the TWPA is less than 0.1 photons.}
    \label{fig-noise-stack}
\end{figure}

Having characterized the overall system noise, next we disentangle its different contributions. 
Input-referred system noise includes contributions from losses in components before the TWPA, the TWPA itself, and components after it.
To evaluate the contributions of the elements preceding the TWPA, we take into account that the noise source is separated from the TWPA input by a bias tee and a directional coupler (see Appendix~\ref{appendix:setup}). 
From their insertion loss $A$ measured separately at \qty{4.2}{\K}, we calculate the transmission efficiency $\eta_1 = 10^{-A / 10}$.
Additionally, considering the noise of the elements after the TWPA, $N_2$, and assuming vacuum noise $N_\mathrm{vac} = 0.5$ at the TWPA input, from Eq.~(\ref{eq:noise_power}) we obtain the total system noise
\begin{equation}
    N_\mathrm{sys} = 2 N_\mathrm{vac} + 2 \frac{1 - \eta_1}{\eta_1}N_\mathrm{vac} + \frac{N_\mathrm{T, exc}}{\eta_1} + \frac{N_2}{\eta_1 G_\mathrm{ss}}
    \label{eq:Nsys}
\end{equation}
for $G_\mathrm{si} / G_\mathrm{ss} = 1$. 
The first term, $2N_\mathrm{vac}=1$ photon, represents the standard quantum limit, while
$N_\mathrm{T,exc}$ captures the TWPA's excess noise above the quantum limit.
The noise of elements following the TWPA, $N_2$, is divided by the TWPA's gain $G_\mathrm{ss}$. 
We measure $N_2$ using the Y-factor method, bypassing the TWPA and considering a single-mode model [Eq.~(\ref{eq:hemt-noise}) in Appendix~\ref{appendix:powercal_from_noise}]. 

All four contributions are depicted in Fig.~\ref{fig-noise-stack} as a stacked plot. 
The total system noise is dominated by the TWPA's excess noise of 0.8 to 1.5 photons, which we attribute to losses in the $\mathrm{SiO_x}$ of the plate capacitors \cite{Macklin2015} and the generation of sidebands \cite{Peng2022}. 
The former could be reduced by using a less dissipative dielectric, such as $\mathrm{SiN_x}$ or hydrogenated a-Si \cite{O’Connell2008, Paik2010, Defrance2024}.
The noise contribution from elements after the TWPA, on average 0.6 photons, could be reduced by decreasing losses between the TWPA and the second amplifier, increasing TWPA gain without increasing its added noise [cf. Fig.~\ref{fig-gain_profile}(f)], or decreasing the added noise of the second amplifier.
In our case, the contribution of the components before the TWPA is small (less than 0.1 photons), while this may not be the case for typical readout chains. 

\section{Conclusion}

We have designed, fabricated, and characterized a Josephson traveling-wave parametric amplifier based on rf-SQUIDs. 
To operate efficiently in the 3WM regime, we have implemented dispersion engineering by multiperiodic capacitance variation, suppressing unwanted mixing processes while maintaining the amplification over a wide frequency range. 
Our device attains an average gain of \qty{20}{\decibel} from \qtyrange[range-units=single]{3.5}{8.5}{\giga \hertz}, thus exceeding one octave bandwidth.
To also characterize saturation power and noise over that same frequency range, we apply wideband techniques based on a calibrated noise source. 
We demonstrate an average saturation power of \qty{-84}{dBm}, one order of magnitude larger than reported for JTWPAs \cite{Macklin2015, Planat2020, Ranadive2022, Perelshtein2022, Qiu2023}. This is achieved for a comparable number of nonlinear elements thanks to the more favorable trade-off between saturation power and device length when using rf-SQUIDs. 
We have explicitly analyzed noise contributions from different circuit components, determining that our total system noise is dominated by the TWPA's excess noise of \numrange[]{0.8}{1.5} photons. 
While this is already on par with existing works, it could be straightforwardly reduced by decreasing the dielectric loss in the parallel plate capacitors.
Notably, we show that the lowest system noise does not coincide with the TWPA's maximum gain, emphasizing that TWPAs should be tuned to minimize the system noise instead of maximizing the gain. 
We also show that the minimum of total system noise aligns with the maximum of signal-to-noise ratio improvement $\Delta \mathrm{SNR}$, validating the latter as a more practical tune-up approach for users. 

In scaling up quantum computing, the high saturation power of our device should allow for either an increase in the number of qubits per readout line or a decrease in intermodulation distortion between frequency-multiplexed qubits through a larger overhead
to the saturation power \cite{Remm2023}.
We believe that our device, with its wide bandwidth and high phase-sensitive extinction ratio of \qty{58}{\decibel}, has great potential for single-mode and two-mode squeezing and the generation of broadband entangled microwave radiation \cite{Grimsmo2017, Esposito2022, Perelshtein2022, Qiu2023}, with applications in quantum sensing and continuous-variable quantum computing. 
We are convinced that rf-SQUID based TWPAs can be further developed to be nonreciprocal devices featuring backward isolation \cite{Ranadive2024, Malnou2024_TWPAC}. 
This would potentially remove the need for bulky, magnetic isolators between qubit devices and the TWPA, which, together with the demonstrated high saturation power, would aid the efforts to scale up superconducting-circuits-based quantum computers.

\begin{acknowledgments}
This project has received support from the European Union’s Horizon Europe 2021-2027 project TruePA (Grant No. 101080152), and from the German Federal Ministry of Education and Research (BMBF) within the framework programme “Quantum technologies – from basic research to market” (Grant No. 13N15949). 
We thank Thomas Weimann, Alexander Fernandez Scarioni, Jan Blohm, Niels Ubbelohde, Judith Felgner, Maikel Petrich, Rolf Gerdau, Bhoomika R. Bhat, and the clean room staff for their assistance with device fabrication. 
We are grateful to Alexander B. Zorin, Marat Khabipov, and Ralf Dolata for their theoretical and experimental input. 
We acknowledge all members of PTB's coherent superconducting quantum circuits group for support with the experimental equipment and helpful discussions. 
We thank Mark Bieler for his constructive feedback on the manuscript.
We are grateful to Linus Andersson, Simone Gasparinetti, and Robert Rehammar for providing low insertion loss IR filters \cite{Rehammar2023}. 

\end{acknowledgments}

\section*{Data availability statement}

The data that support the findings of this article are openly available \cite{Gaydamachenko2025}.

\section*{Author contributions}

V.G. and C.K. contributed equally to this work.
V.G. and C.K. designed the device, measured and analyzed the data, with input from L.G.
V.G. fabricated the device. 
C.K. developed the experimental methods and theoretical models.
V.G. and L.G. wrote the initial manuscript, which was refined by all authors.
L.G. supervised the project.

\appendix

\section{Gain saturation}\label{appendix:saturation}
The effect of pump depletion due to conversion of pump power to signal and idler power has been analyzed analytically for a 3WM-TWPA in the absence of phase mismatch and loss in Ref.~\cite{Zorin2016}. 
The resulting signal power gain, depending on the signal input power $P_\mathrm{s}^\mathrm{in}$, is
\begin{equation}
	G(P_\mathrm{s}^\mathrm{in}) = \frac{1}{\mathrm{dn}^2(gN,\mathrm{k})} ,
	\label{eq:Pssat_dn_formula}
\end{equation}
where $g=(1/4)|\beta|k_\mathrm{p} a \phi_\mathrm{p} \sqrt{1-\delta^2}$ denotes the exponential gain coefficient for a dimensionless detuning $\delta=|f_\mathrm{s}-f_\mathrm{p}/2|/(f_\mathrm{p}/2)$, and with the Jacobi elliptic function $\mathrm{dn}(u,\mathrm{k})$ (see e.g., Ref.~\cite{Olver2010}) for the modulus 
\begin{equation}
	\mathrm{k} = \frac{1}{\sqrt{1 + f_\mathrm{p} P_\mathrm{s}^\mathrm{in} / f_\mathrm{s} P_\mathrm{p}^\mathrm{in} }}
	,
\end{equation}
where we use an upright symbol to distinguish modulus $\mathrm{k}$ from wavenumber $k$.

We propose the following approximation to Eq.~(\ref{eq:Pssat_dn_formula}) to obtain a more illustrative expression for the signal-power dependence of the gain, where we assume large initial gain $G_0=G(P_\mathrm{s}^\mathrm{in}\rightarrow 0) = \cosh^2 gN \gg 1$ and $\mathrm{k} \approx 1$, i.e., $P_\mathrm{s}^\mathrm{in} \ll P_\mathrm{p}^\mathrm{in}$.
According to Ref.~\cite{Olver2010}, in this limit the Jacobi elliptic function can be approximated to
\begin{equation}
	\mathrm{dn}(u,\mathrm{k}) \approx \frac{1}{\cosh(u)} + \frac{\mathrm{k}'^2}{4} \frac{\tanh(u)}{\cosh(u)} \left[ 	\sinh(u)\cosh(u) + u \right] ,
    \label{eq:dn_approx_1}
\end{equation} 
where $\mathrm{k}'\coloneq \sqrt{1-\mathrm{k}^2} $ is the complement modulus of $\mathrm{k}$.
For $\sinh(u)\cosh(u) \gg u$ and $\sinh^2(u) \approx \cosh^2(u)$ we can further simplify Eq.~(\ref{eq:dn_approx_1}) to
\begin{equation}
	\mathrm{dn}(u,\mathrm{k}) \approx \frac{1}{\cosh(u)} \left[ 1 + \frac{\mathrm{k}'^2}{4} \cosh^2(u)\right],
    \label{eq:dn_approx}
\end{equation}
and find, after inserting Eq.~(\ref{eq:dn_approx}) into Eq.~(\ref{eq:Pssat_dn_formula}), and replacing $\cosh^2 gN = G_0$, 
\begin{equation}
	G(P_\mathrm{s}^\mathrm{in}) \approx \frac{G_0}{1+G_0f_\mathrm{p}P_\mathrm{s}/2f_\mathrm{s}P_\mathrm{p} + (G_0f_\mathrm{p}P_\mathrm{s}/4f_\mathrm{s}P_\mathrm{p})^2}
	\label{eq:Pssat_approx2_formula}
\end{equation}
where powers denote input powers, omitting the superscript for brevity.
Neglecting the quadratic term, and considering frequencies in the center of the gain band, $f_\mathrm{p}\approx 2f_\mathrm{s}$, we obtain a simple expression for the gain saturation due to pump depletion,
\begin{equation}
	G(P_\mathrm{s}^\mathrm{in}) \approx \frac{G_0}{1+G_0P_\mathrm{s}/P_\mathrm{p}}.
	\label{eq:Pssat_approx2_formula__}
\end{equation}
With this, the signal input power causing \qty{1}{\decibel} gain compression, $G= 10^{-1/10}G_0$, can be found as
\begin{equation}
	P_{\mathrm{1dB}}^\mathrm{dBm} \approx \frac{P_\mathrm{p}}{4G_0}  =P_\mathrm{p}^\mathrm{dBm}-G_0^\mathrm{dB}-6\;\text{dB}.
	\label{eq:Pssat_1dBCP_Pp-G-6dB}
\end{equation}
An expression similar to Eq.~(\ref{eq:Pssat_approx2_formula__}) was derived in Ref.~\cite{Kylemark2006} for a 4WM optical-fiber TWPA, 
\begin{equation}
	G(P_\mathrm{s}^\mathrm{in})=\frac{G_0}{1+2G_0P_\mathrm{s}/P_\mathrm{p}}.
	\label{eq:Pssat_Kylemark_formula}
\end{equation}
This equation covers pump depletion as well as signal-power-caused SPM and XPM. 
The resulting 1~dB compression point,
\begin{equation}
	P_{\mathrm{1dB}}^\mathrm{dBm} \approx \frac{P_\mathrm{p}}{8G_0}  =P_\mathrm{p}^\mathrm{dBm}-G_0^\mathrm{dB}-9\;\text{dB},
	\label{eq:Pssat_1dBCP_Pp-G-9dB}
\end{equation}
is 3~dB lower than that of a Kerr-free 3WM-TWPA. 
During the drafting of our manuscript, Le Gal et al. published a preprint dedicated to gain compression in 4WM TWPAs \cite{Gal2025}.

\section{Fabrication}
\label{appendix:fab}

The TWPA is fabricated at PTB's clean room center in Nb trilayer technology based on the process outlined in Ref.~\cite{Dolata2005}. 
On a high resistivity ($\rho = \qtyrange[range-phrase = -, range-units = single]{5}{50}{\kilo\ohm\centi\metre}$) 3" intrinsic Si wafer with \qty{200}{\nano \meter} of thermal $\mathrm{SiO_x}$, we deposit \qty{30}{\nano\meter} of $\mathrm{Al_2O_3}$, which serves as an etch stop layer. 
This is followed by \emph{in situ} deposition of a Nb/Al-$\mathrm{AlO_x}$/Nb trilayer using DC magnetron sputtering. 
The base and counter Nb layers have thicknesses of \qtylist[list-units=single]{160;140}{\nano \meter}, while the Al layer is \qty{10}{\nano \meter} thick.
To achieve a critical current density of $j_\mathrm{c} = \qty{130}{\ampere / \centi \meter \squared}$, we oxidize the Al layer at \qty{160}{\milli \bar} of pure oxygen for 32 minutes, including 12 minutes required to reach the target pressure in the chamber. 
After deposition of the trilayer, the wafer is covered with a \qty{50}{\nano \meter} thick $\mathrm{SiO_x}$ deposited using plasma-enhanced chemical vapor deposition (PECVD) to protect Josephson junctions from anodic oxidation later in the process.

The Josephson junctions are patterned using electron beam lithography. 
After development of the negative resist, we perform two sequential dry etching steps: reactive ion etching (RIE) with a $\mathrm{CHF_3}$/Ar mixture to remove the $\mathrm{SiO_x}$ layer, followed by inductively coupled plasma reactive ion etching (ICP-RIE) in an $\mathrm{SF_6}$ plasma to etch Nb. 
The Al-$\mathrm{AlO_x}$ layer is wet etched in a solution of \qty{65}{\percent} $\mathrm{HNO_3}$, \qty{100}{\percent} $\mathrm{CH_3COOH}$, \qty{85}{\percent} $\mathrm{H_3PO_4}$ and $\mathrm{H_2O}$, mixed at a ratio of 1:1:16:2 as described in Ref.~\cite{Krantz2020}.
Following the etching steps, we anodize the base Nb layer.
In the second lithography step, the base Nb electrode is patterned and etched using the same ICP-RIE process as for the counter electrode. 

The entire waver is then covered with \qty{250}{\nano\meter} of PECVD-deposited $\mathrm{SiO_x}$, accounting for a $\sim \qty{10}{\nano \meter}$ loss during Ar pre-cleaning of the wafer before the deposition of the wiring Nb layer.
To enable contact between the Josephson junctions and the wiring layer, we etch windows into the $\mathrm{SiO_x}$ down to the counter electrode of the Josephson junctions. 
These windows are etched using ICP-RIE in a $\mathrm{CHF_3}$ plasma. 

The final step in fabrication is the deposition of the \qty{400}{\nano \meter} thick Nb wiring layer. After the exposure and development of the resist, the wiring layer is patterned using RIE with a $\mathrm{SF_6/O_2}$ mixture at a process pressure of \qty{20}{\Pa} with the substrate temperature maintained at \qty{40}{\degreeCelsius}. These etching parameters help prevent the formation of parasitic conductive structures at the edges of etched features sometimes referred to as fences.

After fabrication, we measure at room temperature the tunnel resistances $R_\mathrm{t}$ of 112 junctions incorporated into a cross-bridge Kelvin resistor (CBKR) structure as a proxy for the critical currents of the Josephson junctions \cite{Berggren1999}. 
This provides insight into the spread of the critical currents, which can be quantified as a coefficient of variation $\sigma(I_\mathrm{c}) / \langle I_\mathrm{c} \rangle$.
For the wafer of the TWPA presented in the main text this variation is unexpectedly large, exceeding 30\%.
Thanks to our design concept, the device performance is robust even against unusually large $I_\mathrm{c}$ spread \cite{Kissling2023}, and still exhibits a gain of \qty{20}{\decibel}.

\section{Measurement setup}
\label{appendix:setup}

\begin{figure}
    \includegraphics[width=8.6cm]{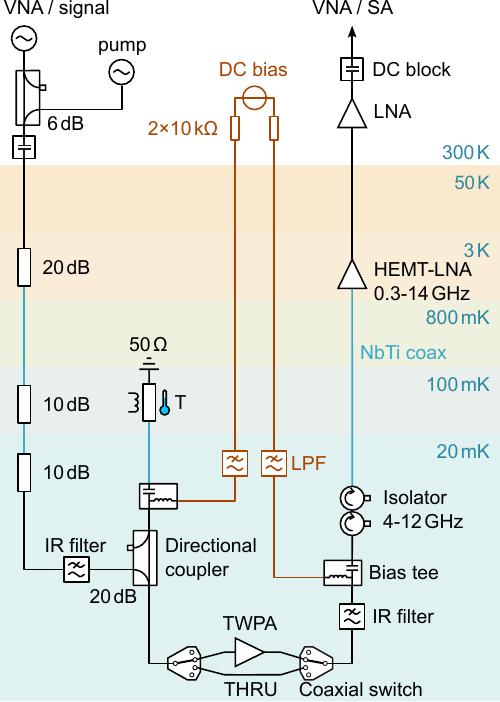}
    \caption{Schematic of the experimental setup. 
    Blue lines denote superconducting NbTi coaxial cables.
    Brown lines denote Thermocoax\textregistered~cables serving as a low pass filter for current biasing of our device.
    The TWPA is surrounded by a mu-metal shield which protects it from external magnetic fields.
    }
    \label{fig-setup}
\end{figure}

In Fig.~\ref{fig-setup} we present a schematic of the experimental setup used in this work. 
The pump tone is coupled to a signal tone at room temperature. 
The signal tone is generated either by a Vector Network Analyzer (VNA) for gain measurements or by a signal generator for SNR measurements.
To avoid ground loops, we use inner/outer DC blocks both at the input and the output as shown in Fig.~\ref{fig-setup}.
The TWPA is protected against photons with sufficient energy for pair breaking with low-insertion-loss infrared filters \cite{Rehammar2023} and a light-tight enclosure.
The dc-current source is connected to the TWPA via bias tees and
Thermocoax\textregistered~cables, the latter of which provide low-pass filtering (LPF) with a cutoff frequency of $\sim \qty{1}{\mega \hertz}$ and a stopband extending to the \unit{\tera \hertz} range \cite{Zorin1995}.
The noise source consists of a \qty{50}{\ohm} load, a heater, and an individually calibrated Ruthenium oxide temperature sensor, thermally anchored to a variable-temperature platform. The load terminates a superconducting coaxial cable made of NbTi, which thermally isolates the noise source from the rest of the setup.
An SP6T coaxial switch allows for \emph{in situ} VNA forward calibration and enables system noise $N_2$ measurements of the chain with the TWPA bypassed (see main text).

\section{Model of the rf-SQUID implementation}
\label{appendix:non-ideal-squid}

\begin{figure}
    \includegraphics[width=8.6cm]{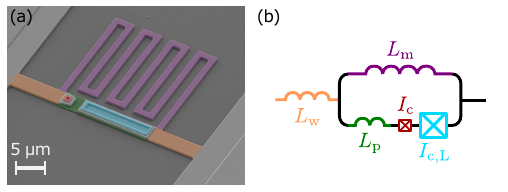}
    \caption{Practical implementation of the rf-SQUID. (a) False-colored scanning scanning electron microscopy image of an rf-SQUID [colors according to panel (b)]. (b) Corresponding circuit schematic, containing wire inductance $L_\mathrm{w}$, meander inductance $L_\mathrm{m}$, parasitic inductance $L_\mathrm{p}$, critical current $I_\mathrm{c}$, critical current of the large junction $I_\mathrm{c, L} \approx 40 I_\mathrm{c}$. The large junction serves as a via connecting the two metal layers.
    }
    \label{fig-real-rf-SQUID}
\end{figure}

To account for deviations between the idealized model [Fig.~\ref{fig-circuit-diagram}(a)] and the practical implementation of the TWPA unit cell, additional circuit elements are introduced to the model.
Figure~\ref{fig-real-rf-SQUID}(a) shows a false-colored SEM image of the rf-SQUID with the corresponding circuit schematic [panel (b)].
The branch with the Josephson junction contains a parasitic linear inductance $L_\mathrm{p}$.
Apart from that, in our implementation we use a Josephson junction as a via, connecting the base Nb with the counter Nb layer.
This junctions is 40 times larger than the Josephson junction with critical current $I_\mathrm{c}$. 
While the latter dominates the nonlinear behavior of the rf-SQUID, the former merely adds quasi-linear inductance $L_{\mathrm{J}0, \mathrm{L}} = \varphi_0 / I_\mathrm{c, L}$, with $I_\mathrm{c, L} \approx 40 I_\mathrm{c}$.
All the inductances contribute to the loop inductance, modifying the formula for the screening parameter,
\begin{equation}
\beta_L = \frac{I_\mathrm{c} L_\mathrm{loop}}{\varphi_0} = \frac{(L_\mathrm{m} + L_\mathrm{p} + L_{\mathrm{J}0, \mathrm{L}})}{\varphi_0}.
\end{equation}
and the total rf-SQUID inductance,
\begin{equation}
L_\mathrm{SQ} = L_\mathrm{m} \frac{1 + \alpha_\mathrm{p} \beta_{L} \cos \phi_\mathrm{dc}}{1 + \beta_{L} \cos \phi_\mathrm{dc}}, \label{eq:L_squid_modified}
\end{equation}
introducing coefficient $\alpha_\mathrm{p} = (L_\mathrm{p} + L_{\mathrm{J}0, \mathrm{L}})/(L_\mathrm{p} + L_{\mathrm{J}0, \mathrm{L}} + L_\mathrm{m})$, describing the ratio of parasitic inductance to the total loop inductance.
Further, two neighboring rf-SQUIDs are connected to each other with a wire having the inductance $L_\mathrm{w}$, contributing to the unit cell inductance $L_\mathrm{cell} = L_\mathrm{w} + L_\mathrm{SQ}$. 

Fitting $L_\mathrm{cell}(\phi_\mathrm{dc})$ to experimental data [shown in Fig.\ref{fig-s21-vs-power}(d)] yields the values $L_\mathrm{m}=\qty{58.6}{\pH}$, $L_\mathrm{w}=\qty{37.0}{\pH}$, $L_\mathrm{p} + L_{\mathrm{J}0, \mathrm{L}}=\qty{8.9}{\pH}$, and $I_\mathrm{c}=\qty{0.93}{\micro\ampere}$. 
Furthermore, we determined the specific capacitance by measurements on test structures at \qty{4.2}{\K}. 
As a result, at $\phi_\mathrm{dc}=\pi/2$ the characteristic impedance of the TWPA transmission line is \qty{52}{\ohm}. 

See Ref.~\cite{Kissling_PhD_2025} for further details of the model for the rf-SQUID implementation used here.

\begin{figure*}
    \includegraphics[width=17.8cm]{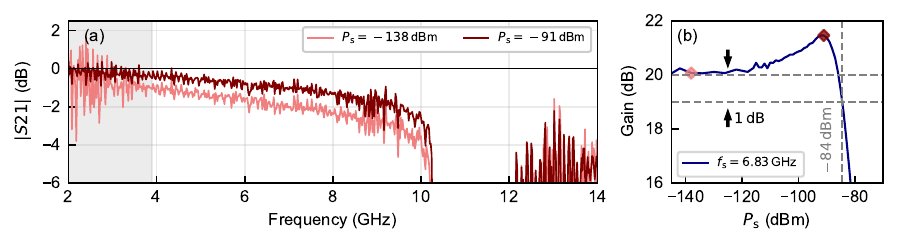}
    \caption{Linear response of the TWPA and gain dependence on the signal input power. (a) $S_{21}$ transmission of the unpumped TWPA at 20 mK for two different signal powers. The gray shaded region indicates frequencies where the isolator is out of band.
    The transmission dip at frequencies from \qtyrange[range-units = single]{10.2}{12}{\giga\hertz} is the designed stopband.
    We attribute the loss to two-level states (TLS) \cite{McRae2020}, which saturate at higher powers.
    (b) Power gain as a function of the input signal power at signal frequency of \qty{6.83}{\giga \hertz}. 
    We attribute the increase of gain values to a gradual saturation of TLS loss. 
    The two diamonds correspond to the signal powers shown in panel (a) with same colors. 
    Throughout the paper and, in particular, for the definition of the 1-dB compression point, we consider the gain at low powers.}
    \label{fig-linear-response}
\end{figure*}

\section{Power-dependent inductance of the rf-SQUID}
\label{appendix:L_cell_vs_power}
As the power of propagating waves in the TWPA increases, the rf-SQUID behavior deviates from the small-signal case. 
This manifests in the $S_{21}$ spectra [see Fig.~\ref{fig-s21-vs-power}(a-c)] as power-dependent frequency shifts of the stopband due to the power dependence of the rf-SQUID inductance. We use this effect for power calibration by aligning experimental with simulation data.

\subsection{Experiment}

To extract the rf-SQUID inductance from experimental data, we measure the phase of $S_{21}$ at \qty{5.5}{\giga \hertz}, where the second harmonic generation is suppressed due to the stopband, which would otherwise cause an additional phase shift. From the phase data, we calculate the wavenumber
\begin{equation}
    k = - \frac{\mathrm{arg}(S_{21})}{a N},
\end{equation}
with the number of cells $N$, assuming $|S_{11}|\ll |S_{21}|$. When the power is increased from $P_0$ to $P$, the wavenumber changes
by
\begin{equation}
    \delta k = - \frac{\mathrm{arg}(S_{21}(P)) -\mathrm{arg}(S_{21}(P_0))}{aN}.
\end{equation}
The cell inductance $L_\mathrm{cell}(P_0)$ is extracted from the fit of a CTM model to $S_{21}$ [see Fig.~\ref{fig-s21-vs-power}(e)].
At frequencies far below the first gap the chromatic dispersion of the TWPA is small, so $k$ is approximately linear and can be expressed via the phase velocity $v_\mathrm{ph}$ as
\begin{equation}
    k_\mathrm{lin}(P) = \frac{\omega}{v_\mathrm{ph}}=\omega a^{-1} \sqrt{L_\mathrm{cell}(P) \bar C_\mathrm{g}}, \label{eq:k_from_p}
\end{equation}
where $\bar C_\mathrm{g}$ denotes the average ground capacitance. 
Rearranging Eq.~(\ref{eq:k_from_p}) and inserting $k_\mathrm{lin}(P) = k_\mathrm{lin}(P_0) + \delta k$ then yields
\begin{equation}
    L_\mathrm{cell}(P)= \frac{1}{\bar C_\mathrm{g}} \left( \frac{k_\mathrm{lin}(P_0) + \delta k (P)}{\omega a^{-1}}  \right)^2.
\end{equation}

\subsection{Simulation}

To obtain the power-dependent inductance of the rf-SQUID model shown in Fig.~\ref{fig-real-rf-SQUID}(b), we carry out WRspice simulations of the model, using the circuit parameters from Appendix~\ref{appendix:non-ideal-squid}. 
The impedance of the rf-SQUID, biased at $I_\mathrm{dc}$ and driven by a complex ac current amplitude $I_{\mathrm{ac}}$, can be found as
\begin{equation}
    Z_{\mathrm{cell}} = R_{\mathrm{cell}} + i \omega L_{\mathrm{cell}} =  V_{\mathrm{SQ}} / I_{\mathrm{ac}}, \label{eq:impedance-wrspice}
\end{equation}
where $V_{\mathrm{SQ}}$ is the complex voltage drop across the SQUID. The resulting SQUID inductance is
\begin{equation}
    L_{\mathrm{cell}}(I_{\mathrm{ac}}) = \frac{\operatorname{Im}Z_{\mathrm{cell}}(I_{\mathrm{ac}})}{\omega}.
\end{equation}
In the TWPA, the current $I_{\mathrm{ac}}$ flowing through the rf-SQUID corresponds to an incident power $P\approx 0.5 |I_{\mathrm{ac}}|^2 Z(\phi_\mathrm{dc}) \approx 0.5 |I_{\mathrm{ac}}|^2 Z_0$ to the TWPA for frequencies much lower than the cutoff frequency of the TWPA transmission line \cite{Kissling_PhD_2025}. 

As stated in the main text, using this method we extract an input line attenuation of \qty{69}{\decibel}, which is \qty{3}{\decibel} more than from the power calibration using the HEMT-LNA noise measurement (see Appendix~\ref{appendix:powercal_from_noise}). 
We attribute the observed discrepancy to the fact that the experimentally determined rf-SQUID inductance $L_\mathrm{cell}(P)$ should be seen as an average inductance of all unit cells of the TWPA. Moreover, accounting for insertion loss of the TWPA, the input power is higher than the average power inside the TWPA. 
Ignoring the insertion loss therefore causes an underestimation of the input power, or an overestimation of the attenuation of the input line. 
Further, we note that the experimental extraction of the rf-SQUID circuit parameters involves two fits, which are additional sources of uncertainty. 

\section{Power-dependent transmission loss}
\label{appendix:tls}

Figure~\ref{fig-linear-response}(a) shows the normalized $S_{21}$ transmission of the TWPA, biased at $\Phi_\mathrm{ext} \approx 0.33 \Phi_0$ without a pump, measured at two different probe powers.
The insertion loss of the device decreases by up to \qty{1.5}{\decibel} at \qty{9.8}{\giga \hertz} when increasing the probe power to \qty{-91}{dBm}.
We attribute this effect to the saturation of two-level-state (TLS) loss \cite{O’Connell2008, McRae2020}.
This power-dependent change of insertion loss impacts gain measurements and must be accounted for.

Figure~\ref{fig-linear-response}(b) shows a typical saturation power measurement.
Before the gain begins to drop due to pump depletion, we observe an increase by \qtyrange[range-units=single, range-phrase=--]{1}{2}{\decibel}, an effect which has been observed by other groups and is attributed to TLS saturation, among other effects \cite{Planat2020}.
For typical qubit readout applications powers are often below \qty{-120}{dBm}.
To account for such applications, all gains reported in this paper are measured at powers below \qty{-120}{dBm}.
Similarly, for saturation power measurements, we take the gain value in the flat, low-power regime as a reference.

Note that we use the conventional definition for gain saturation as the signal power at which the gain decreases by \qty{1}{\decibel}. 
However, the maximal signal power for which the amplifier operates linearly (e.g., the gain is constant within $\pm \qty{1}{\decibel}$) can be smaller than the saturation power, when the gain increases by more than \qty{1}{\decibel} before it drops.

\section{Gain ripple}
\label{appendix:effect-on-ripples}
\sisetup{uncertainty-mode = separate}

\begin{figure}[b]
    \includegraphics[width=8.6cm]{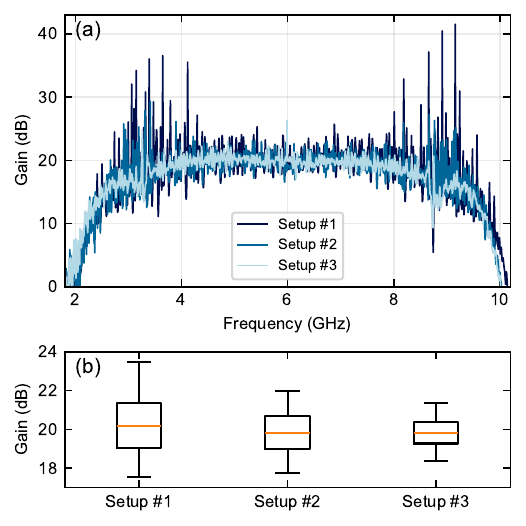}
    \caption{
    Gain profiles of the TWPA using different peripheral components. 
    (a) Each gain profile is obtained after tuning the TWPA to have the smallest gain ripple, while having an average gain of \qty{20}{\decibel}.
    (b) Boxplots showing amplifier gain distribution for frequencies from \qtyrange[range-units=single]{4}{8}{\giga \hertz} for different setups. Whiskers (boxes) represent the 5th (25th) and 95th (75th) percentiles. Orange line indicates the median value. 
    }
    \label{fig-twpa-ripples}
\end{figure}

To reduce gain ripple, reflections inside and outside the TWPA have to be decreased.
These reflections originate from multiple impedance mismatches, including the TWPA transmission line itself, stopband sidelobes resulting from our dispersion engineering technique \cite{Gaydamachenko2022}, TWPA packaging elements such as bond wires, and external circuitry.
It is crucial to reduce reflections not only within the signal band, but also at the pump frequency. 
Reflections at the pump frequency lead to backward gain, which further increases gain ripple.
Since pump frequencies are at the edge of the rated bandwidth of the isolator in our setup, reflections of the pump wave are expected. 

Figure~\ref{fig-twpa-ripples}(a) shows the TWPA gain profile measured in three cooldowns with different experimental setups. 
Each setup variant has different components that influence ripples of the resulting gain profile. 
In all cases, the TWPA is tuned to minimize gain ripple while providing an average gain of \qty{20}{\decibel}. 
The boxplots in Fig.~\ref{fig-twpa-ripples}(b) illustrate the distribution of the gain values across the \qtyrange[range-units=single]{4}{8}{\giga \hertz} band. 
In Setup \#1 \qty{90}{\percent} of the gain values fall within the range $20.3^{+3.2}_{-2.7}$\,dB.
In Setup \#2 the SMA bulkheads of the TWPA enclosure were replaced with better impedance-matched components, and this setup is used for the results reported in the main text. 
Here, \qty{90}{\percent} of the gain values fall within the range \qty[separate-uncertainty-units = bracket]{19.8(2.4)}{\decibel}.
To investigate the effect of reduced reflections, we use a 10 dB attenuator at the TWPA output in Setup \#3.
This modification reduced the spread of gain values to \qty[separate-uncertainty-units = bracket]{19.8(1.5)}{\decibel}. 
Although this setup is not suitable for readout applications due to degraded SNR, it shows that using better-matched external circuitry has the potential of significantly reducing gain ripple. The same applies to components before the TWPA.

\section{Phase-sensitive parametric amplification}
\label{app:phase-sensitive-gain}

A special case in parametric amplification occurs when the signal and idler frequencies coincide, $f_\mathrm{s}=f_\mathrm{i}$, also called degenerate parametric amplification. 
In this regime, parametric amplification is phase-sensitive, i.e., the signal gain depends on the phase of the signal wave relative to that of the pump wave, while for other frequencies the gain is independent of phase (phase-preserving). 
Phase-sensitive gain enables interesting effects such as single-mode squeezing and amplification with noise below the standard quantum limit \cite{Perelshtein2022, Qiu2023}.

In a 3WM TWPA phase-sensitive parametric amplification occurs at $f_\mathrm{s}=f_\mathrm{i}=f_\mathrm{p} / 2$, in the center of the gain profile. This is in contrast to (single-pump) 4WM TWPAs, where it coincides with the pump frequency, $f_\mathrm{s}=f_\mathrm{i}=f_\mathrm{p}$, and is therefore not accessible. 
To measure the dependence of the gain on the signal phase, we vary the phase of the signal generator relative to that of the pump generator. The signal and pump generators and the spectrum analyzer are synchronized. 
The phase-sensitive gain is measured at $f_\mathrm{s}=f_\mathrm{p}/2$, while the phase-preserving gain is measured at a frequency detuned by \qty{10}{\kilo \hertz} from $f_\mathrm{p}/2$. 

\begin{figure}
    \includegraphics[width=8.6cm]{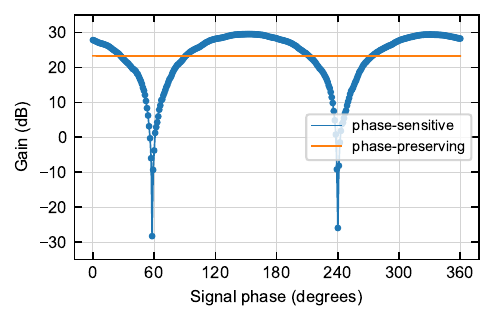}
    \caption{Phase-sensitive and phase-preserving parametric amplification. 
    The blue line shows the signal gain measured at $f_\mathrm{s}=f_\mathrm{i}=f_\mathrm{p}/2$, while for the orange line the signal frequency is detuned by \qty{10}{\kilo \hertz} from that condition. The phase of the signal generator is varied relative to that of the pump generator. 
    The phase-sensitive gain ranges across \qty{58}{\decibel} between maximum amplification and maximum deamplification. 
    The maximum phase-sensitive gain is ca. \qty{6}{\decibel} higher than the phase-preserving gain. 
    Both gains are measured as the ratio of the signal transmission with the pump on and off.
    }
    \label{fig-phasesensitive}
\end{figure}

Figure~\ref{fig-phasesensitive} shows the phase-sensitive signal gain in comparison to the phase-preserving gain, both in dependence on the signal phase. 
While the phase-preserving gain does not show any dependence on the signal phase, the phase-sensitive gain varies periodically between amplification and deamplification. The maximum amplification is ca. \qty{6}{\decibel} above the phase-preserving gain due to the constructive interference of equal-amplitude signal and idler waves, and the maximum deamplification reaches more than \qty{25}{\decibel}. 
The ratio of the two, also referred to as the phase-sensitive extinction ratio, is \qty{58}{\decibel}, which compares to \qty{48}{\decibel} reported for a resonator-based Josephson parametric amplifier \cite{Zhong2013} and \qty{56}{\decibel} demonstrated with the dual-pump 4WM TWPA reported in Ref.~\cite{Qiu2023}. 

We further observe pump subharmonic generation, as a peak at $f_\mathrm{p}/2$ present in the output spectrum, when no signal is applied to the TWPA. This effect is also known as half-tone generation or oscillation period doubling \cite{Zorin2011, Khabipov2022}. Subharmonic generation of the pump was not observed for the 4WM operation points shown in Fig.~\ref{fig-4wm}. 

\section{Four-wave mixing regime}

\begin{figure}
    \label{appendix:4wm}
    \includegraphics[width=8.6cm]{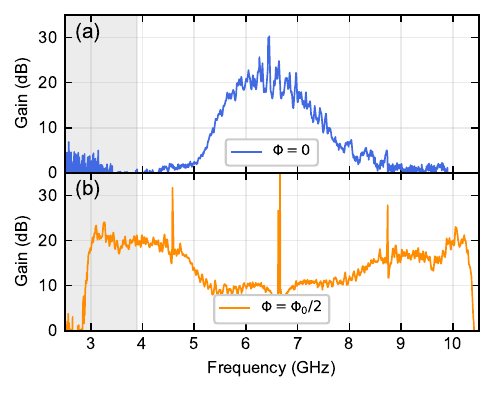}
    \caption{4WM gain profiles of the TWPA. 
    (a) At a flux bias of $\Phi_\mathrm{ext}=0$ the second-order nonlinearity vanishes and Kerr nonlinearity is utilized for amplification. With the pump at $f_\mathrm{p}=\qty{10.56}{\giga \hertz}$ and $P_\mathrm{p} \approx \qty{-54}{dBm}$, phase-matching is achieved around \qty{6.5}{\giga \hertz} and the device provides gain of ca. \qty{20}{\decibel}. A second gain lobe is located around \qty{17}{\giga \hertz} (not shown). 
    (b) At a flux bias of $\Phi_\mathrm{ext}=\Phi_0 / 2$ the Kerr-nonlinearity coefficient is negative and the gain profile has an M-shape similar to Ref.~\cite{Ranadive2022}, the pump parameters are $f_\mathrm{p} = \qty{6.66}{\giga \hertz}$ and $P_\mathrm{p} \approx \qty{-56}{dBm}$. 
    Two sharp peaks are caused by a resonance in the sample package at \qty{8.8}{\giga \hertz}, 
    which additionally creates an image peak at \qty{4.52}{\giga \hertz}.}
    \label{fig-4wm}
\end{figure}

Our rf-SQUID-based TWPA can also be operated in the 4WM regime by biasing it accordingly. 
In this case, amplification is enabled by the Kerr nonlinearity and is governed by the phase-matching condition $2 k_\mathrm{p} = k_\mathrm{s} + k_\mathrm{i}$ along with the energy conservation law $2 f_\mathrm{p} = f_\mathrm{s} + f_\mathrm{i}$.
Taking into account rf-SQUID inductance values of ca. \qtyrange[range-units=single, range-phrase=--]{86}{108}{\pH} and an average capacitance of \qty{36}{\fF} determined from test structure measurements, we obtain an average characteristic impedance ranging from \qtyrange[range-units=single]{49}{55}{\ohm}. Therefore, the low flux tunability of the inductance leads to reasonable impedance matching for all flux bias points, probably limited by the sidelobes of the stopband \cite{Gaydamachenko2022} and not by the flux-dependent average impedance of the transmission line.

Figure~\ref{fig-4wm} shows the gain profiles at two flux-bias points, where the Kerr nonlinearity is maximized each.
In Fig~\ref{fig-4wm}(a) the flux bias is $\Phi_\mathrm{ext}=0$ and the corresponding Kerr-nonlinearity coefficient $\gamma$ is positive. 
Setting the pump to \qty{10.56}{\giga \hertz}, close to the lower edge of the stopband, phase-matching is achieved, similarly to works in Refs.~\cite{HoEom2012,  Macklin2015, White2015, Planat2020}. 
The device exhibits an average gain of \qty{20}{\decibel} from \qtyrange[range-units=single]{6}{7}{\giga \hertz}. 
A second gain band lies at ca. \qty{17}{\giga \hertz}, which is outside of the isolator bandwidth and is not measured. 

Figure~\ref{fig-4wm}(b) shows the gain profile at the flux bias $\Phi_\mathrm{ext}=\Phi_0 / 2$ where the Kerr-nonlinearity coefficient is negative. 
Phase-matching can be achieved when the phase shifts caused by SPM and XPM compensate the chromatic dispersion, as utilized in Ref.~\cite{Ranadive2022}. 
An M-shape gain profile is observed. The positions of the two gain lobes can be tuned via the pump frequency across several GHz. 

The measurements have been performed with a \qty{10}{\decibel} attenuator at the TWPA output to avoid potential reflections caused by the impedance mismatch (cf. Appendix~\ref{appendix:effect-on-ripples}). 
For proper calibration the through line used for calibration was also equipped with a \qty{10}{\decibel} attenuator of the same type.

\section{Wideband power calibration}
\label{appendix:powercal_from_noise}

\begin{figure*}
    \includegraphics[width=14.24cm]{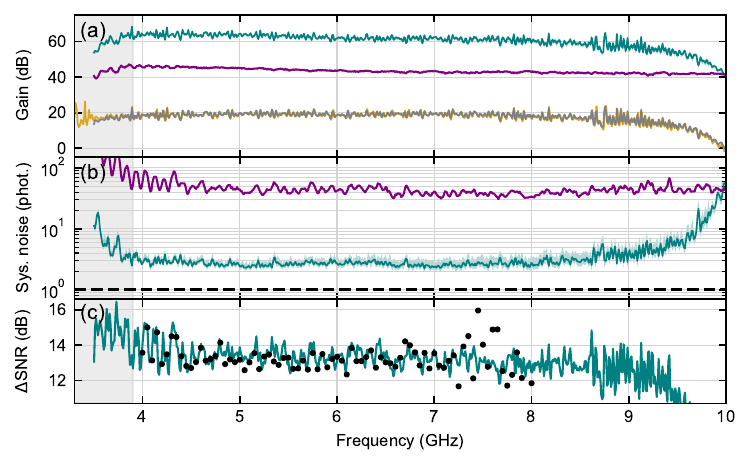}
    \caption{System gain, system noise, and SNR-improvement. 
    (a) Teal and purple curves denote the system gain of the amplifier chains including the TWPA ($G_\mathrm{sys}$) and bypassing it ($G_2$), respectively. 
    The difference between the two curves results in the TWPA gain profile (gray curve), which agrees with the one measured with the VNA, showing an RMS error of \qty{0.6}{\decibel} (gold curve, cf. Fig.~\ref{fig-gain_profile}). 
    (b) Comparison of the total system noise for two amplifier chains: one including the TWPA (teal) and one bypassing it (purple). 
    (c) Teal curve shows $\Delta \mathrm{SNR}$ analytically obtained from the total system noise measurements. Scatter dots indicate directly measured $\Delta \mathrm{SNR}$.
    From \qtyrange[range-units=single]{4}{7}{\giga \hertz} the RMS error equals \qty{0.5}{\decibel}, relative error of $\approx \qty{4}{\percent}$.
    The deviation between the measurements and analytical predictions above \qty{7}{\giga \hertz} is under investigation.}
    \label{fig-Gsys_Nsys_SNR}
\end{figure*}

Accurately determining the saturation power of parametric amplifiers operating at mK-temperatures is a challenging task and requires power calibration at a cryogenic reference plane. 
To address this challenge, we leverage the noise measurements of the amplifier chain without the TWPA, where the HEMT-LNA is the first amplifier in the chain. 
The single-mode model of the output noise of the amplification chain is
\begin{equation}
    N_\mathrm{out}(T) =  \eta_1 G_2 \left( N_\mathrm{in}(T) + \frac{(1 - \eta_1)}{\eta_1} N_\mathrm{vac} + \frac{N_2}{\eta_1}\right). \label{eq:hemt-noise}
\end{equation}
Fitting Eq.~(\ref{eq:hemt-noise}) to the experimental data we obtain the gain $G_2$ between the output of the TWPA and the output of the chain over a broad frequency range [purple curve in Fig.~\ref{fig-Gsys_Nsys_SNR}(a)]. 
Importantly, to correctly measure the noise power with the SA, an RMS detector must be used, as other detector settings may introduce offsets in the displayed noise power level. While these offsets do not affect the fitting result $N_2$, they lead to an incorrect system gain $G_2$ by the same offset.

Additionally, we measure the roundtrip transmission $S_{21, \,\mathrm{roundtrip}}$ of the setup in the same frequency range. For this measurement the VNA is calibrated to have a reference plane at its own ports. Combining these two measurements, we can estimate the insertion loss of the input line as
\begin{equation}
    IL^{\mathrm{dB}} = G_2^{\mathrm{dB}} - S_{21, \,\mathrm{roundtrip}}^{\mathrm{dB}}
\end{equation}
across the whole signal frequency range. Repeating this procedure in several cooldowns and with different HEMT-LNAs we find the extracted input line insertion loss consistent within less than \qty{2}{\decibel}.

\section{Y-factor method and SNR improvement}
\label{appndix:snr-from-y-factor}

In this section, we demonstrate that the $\Delta \mathrm{SNR}$, measured directly, can be independently reproduced using the results of the system noise measurements obtained with the modified Y-factor method. 
This serves as a consistency check for our noise characterization. 

\subsection{Y-factor}
Figure~\ref{fig-Gsys_Nsys_SNR}(a) and (b) show the system gain and noise of the TWPA-based amplifier chain, $G_\mathrm{sys}$ and $N_\mathrm{sys}$ (teal curves), as well as the gain and noise of the elements after the TWPA, $G_2$ and $N_2$ (purple curves). 
From these quantities the TWPA gain can be calculated as $G_\mathrm{ss} = G_\mathrm{sys} / \eta_1  G_2$, shown in Figure~\ref{fig-Gsys_Nsys_SNR}(a) as a gray curve. It is consistent with the corresponding VNA measurement of the TWPA gain (gold curve) within an RMS error of \qty{0.6}{\decibel} in the \qtyrange[range-units=single]{4}{8}{\giga \hertz} range. 

\subsection{SNR improvement}
To calculate $\Delta \mathrm{SNR}$ at a frequency $f$, one needs to know the TWPA's added noise $N_\mathrm{T}=N_\mathrm{vac} + N_\mathrm{T,exc}$, power gain $G_\mathrm{ss}$, transmission efficiency $\eta_\mathrm{T}=10^{\,S_{21}^\mathrm{dB}/10}$ (which accounts for signal attenuation by the TWPA when the pump is off), and the total system noise of the chain following the TWPA, $N_2$.
With these quantities, the SNR improvement can be found as
\begin{equation}
    \Delta \mathrm{SNR} = \frac{N_\mathrm{in} + \frac{1 - \eta_\mathrm{T}}{\eta_\mathrm{T}}N_\mathrm{vac}+\frac{N_\mathrm{2}}{\eta_\mathrm{T}}}{N_\mathrm{in} + N_\mathrm{T} + \frac{N_\mathrm{2}}{G_\mathrm{ss}}}.
    \label{eq:snr-an}
\end{equation}

Figure~\ref{fig-Gsys_Nsys_SNR}(c) compares $\Delta \mathrm{SNR}$ obtained using two different methods. 
Teal curve represents the calculated $\Delta \mathrm{SNR}$ using Eq.~(\ref{eq:snr-an}) with the results of the system noise measurements.
The scatter dots show $\Delta \mathrm{SNR}$ obtained with the SA.
To obtain SNR, we send a probe tone from the signal generator at an incident power of $\approx \qty{-124}{dBm}$ and measure the output spectrum with and without the pump. 
The parameters of the SA are: a resolution bandwidth of \qty{100}{\hertz}, a video bandwidth of \qty{1}{\hertz}, and a span of \qty{20}{\kilo \hertz}.
For the signal power we take the maximum value in the measured span, while the noise power is given by the average of points outside the peak. 
SNR improvement is defined as $\Delta \mathrm{SNR} = \mathrm{SNR}_\mathrm{pump\,on} / \mathrm{SNR}_\mathrm{pump\,off}$

From \qtyrange[range-units=single]{4}{7}{\giga \hertz} we find an RMS error of \qty{0.5}{\decibel}, reinforcing the reliability of our measurement. We emphasize that it is not possible to obtain TWPA added noise from SNR measurements alone without additional measurements.

\section{Sideband-induced excess noise}\label{appendix:sb_excess_noise}

Due to its broadband behavior, a TWPA converts not only the input noise at the idler frequency but also the noise at higher sideband frequencies into output signal noise, ultimately leading to an added noise above the quantum limit. 
Furthermore, noise incident on the output of the TWPA can convert into output signal noise, either via parametric forward-backward coupling, or via reflections and subsequent parametric forward-forward coupling \cite{Peng2022}.
To quantify these contributions to the added noise of the TWPA, we adopt the approach outlined in Ref.~\cite{Peng2022a}, which we implement using WRspice  simulations of a circuit model of the TWPA. 

The input noise $N_\mathrm{in}^\mathrm{m}$ at frequency $f_\mathrm{m}$, $\mathrm{m}\in\{\mathrm{s,i,p\!+\!s,p\!+\!i,2p\!+\!s,...}\}$, is converted to signal output noise $N_\mathrm{out}^\mathrm{s} = G_\mathrm{s,m} N_\mathrm{in}^\mathrm{m}$ via the corresponding photon-conversion gain $G_\mathrm{s,m}$. 
Similarly, noise at frequency $f_\mathrm{m}$ incident to the output of the TWPA is converted to signal output noise via the backward-forward photon-conversion gain $G_\mathrm{s,m}^\leftrightarrows$. 
To obtain these conversion gains, the input (output) is driven at frequency $f_\mathrm{m}$, and the output response at $f_\mathrm{s}$ is measured, 
\begin{align}
    G_\mathrm{s,m} &= \frac{|b_\mathrm{2,s}|^2}{|a_\mathrm{1,m}|^2} \frac{f_\mathrm{m}}{f_\mathrm{s}}, \\
    G_\mathrm{s,m}^\leftrightarrows &= \frac{|b_\mathrm{2,s}|^2}{|a_\mathrm{2,m}|^2} \frac{f_\mathrm{m}}{f_\mathrm{s}},
\end{align}
where $a_\mathrm{q,j}$ and $b_\mathrm{q,j}$ are the incident and outgoing power-wave quantities at port $\mathrm{q}$ and frequency $f_\mathrm{j}$. The coefficient $f_\mathrm{m}/f_\mathrm{s}$ accounts for different photon energies. 
With the noise $N_\mathrm{in,q}^\mathrm{m}$ incident to the TWPA at port $\mathrm{q}$, and the contribution of internal loss, $N_\mathrm{exc,loss}$, the output signal noise is 
\begin{align}
    N_\mathrm{out}^\mathrm{s} &= \sum_{m} G_\mathrm{s,m} N_\mathrm{in,1}^\mathrm{m} 
    + \sum_{m} G_\mathrm{s,m}^\leftrightarrows N_\mathrm{in,2}^\mathrm{m}
    + G_\mathrm{s,s} N_\mathrm{exc,loss} \\
    &= G_\mathrm{s,s}\left( N_\mathrm{in}^\mathrm{s} +  \frac{G_\mathrm{s,i}}{G_\mathrm{s,s}} N_\mathrm{in}^\mathrm{i} + N_\mathrm{exc,sideband} + N_\mathrm{exc,loss}\right),
\end{align}
[c.f. Eqs.~(\ref{eq:noise_power}) and (\ref{eq:Nsys}), where $N_\mathrm{T,exc} = N_\mathrm{exc,sideband} + N_\mathrm{exc,loss}$]. 
Considering a well-thermalized environment of the TWPA, and $k_\mathrm{B}T\ll h f_\mathrm{m}$, the noise incident to the TWPA port $\mathrm{q}$ is $N_\mathrm{in,q}^\mathrm{m} = 1/2$ photon. 
Then, the sideband-induced excess noise (including forward-backward coupling) is
\begin{equation}
    N_\mathrm{exc,sideband} = \frac{1}{2}\left(
    \sum_{\substack{m\in \\ \{p+s, p+i, 2p+s,...\}}}  G_\mathrm{s,m}  
    + \sum_{\substack{m\in \\ \{s, i, p+s,...\}}}  G_\mathrm{s,m}^\leftrightarrows  \right).
\end{equation}

\begin{figure}
    \includegraphics[width=8.6cm]{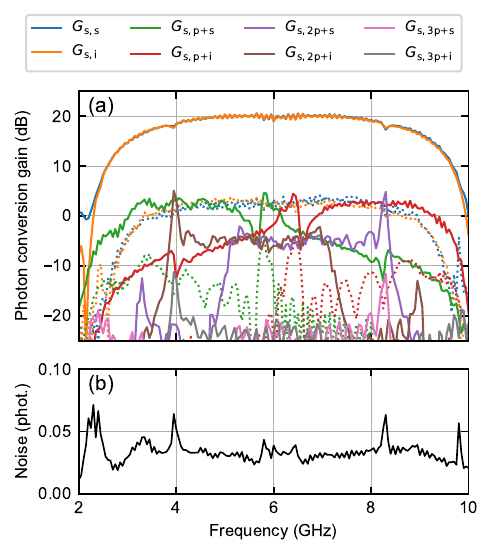}
    \caption{Simulated conversion gains and sideband-induced added noise. 
    (a) Photon-energy-normalized conversion gains quantifying the conversion of incident noise at frequency $f_\mathrm{m}$ ($\mathrm{m}\in\{\mathrm{s,i,p\!+\!s,...}\}$) to signal output noise at $f_\mathrm{s}$. 
    Solid (dotted) lines indicate forward-forward (backward-forward) coupling, i.e., the tone $f_\mathrm{m}$ is incident to the input (output) of the TWPA. 
    (b) Contribution of the sidebands to the added noise of the TWPA, comprising noise from forward-forward-coupled higher sideband frequencies $f_\mathrm{p+s}, f_\mathrm{p+i}, f_\mathrm{2p+s},...$ as well as from all backward-forward-couplings, up to third order of the pump, and considering vacuum noise incident to both the input and output of the TWPA.}
    \label{fig-sb_excess_noise}
\end{figure}

Figure~\ref{fig-sb_excess_noise}(a) shows the simulated photon-conversion gains up to third order. 
The simulation suggests a gain asymmetry $G_\mathrm{si} / G_\mathrm{ss} < \qty{\pm 0.3}{\decibel}$  in the range \qtyrange[range-units=single]{3}{9}{\giga \hertz}. 
Remarkably, all conversion gains of higher sidebands and backward-forward-coupled modes are at least \qtyrange[range-units=single]{10}{15}{\decibel} below the signal power gain $G_\mathrm{ss}$, indicating that they are strongly suppressed due to the multiperiodic capacitance variation.
The corresponding sideband excess noise $N_\mathrm{exc,sideband}$ [Fig.~\ref{fig-sb_excess_noise}(b)] remains well below 0.1 photon across the frequency range. 
This suggests that the contributions of sideband generation and backward-forward coupling to the added noise of the TWPA are rather small compared to those from internal loss. 
The peaks in the sideband-induced excess noise, accompanied by slight dips in the signal gain, can be attributed to favorable sideband generation conditions in the vicinity of higher-order stopbands. Although this effect is not very pronounced, it is undesirable and could, in principle, be mitigated by using a sinusoidal rather than a discrete multiperiodic capacitance profile.

\bibliography{twpa_ge_references}
\end{document}